%Paper: hep-ph/9310364
%From: KAPLAN@HUHEPL.HARVARD.EDU
%Date: Wed, 27 Oct 1993 14:49:22 -0400 (EDT)

\input harvmac
\pretolerance=750
\def\prl#1#2#3{Phys. Rev. Lett. {\bf #1}  (#2) #3}
\def\npb#1#2#3{Nucl. Phys. {\bf B#1}  (#2) #3}
\def\prd#1#2#3{Phys. Rev. {\bf D#1}  (#2) #3}
\def\plb#1#2#3{Phys. Lett. {\bf B#1}  (#2) #3}
\def\sss{\scriptscriptstyle}
\def\pd{\partial}
\def\dg{\dagger}
\Title{\vbox{\hbox{HUTP-93/A030}}} {\vbox{
\hbox{\centerline{Nonperturbative Matching for Field}}\medskip
\hbox{\centerline{Theories with Heavy Fermions$^*$}}
}}\footnote{}{$^*$Research
supported in part by the National Science Foundation,
under grant \# PHY-9218167, and in part by the Texas National
Research Laboratory Commission under grant \# RGFY93-278B.}
\centerline{Howard Georgi}
\medskip
\centerline{Lev Kaplan}
\medskip
\centerline{David Morin}
\bigskip
\centerline{Lyman Laboratory of Physics}
\centerline{Harvard University}
\centerline{Cambridge, MA 02138}

\vskip .3in
We examine a paradox, suggested by Banks and Dabholkar, concerning
nonperturbative effects in an effective field theory which is obtained by
integrating out a generation of heavy fermions, where the heavy fermion
masses arise from Yukawa couplings. They argue that light fermions
in the effective theory appear to decay via
instanton processes, whereas their decay is forbidden in the full theory. We
resolve this paradox by showing that such processes in fact do not occur in
the effective theory, due to matching corrections which cause the relevant
light field configurations to have infinite action.

\Date{10/93}

Some time ago, D'Hoker and Farhi \ref\df{E. D'Hoker and E. Farhi,
\npb{248}{1984}{59}; \npb{248}{1984}{77}.} studied the decoupling of a heavy
fermion which gets its mass through a Yukawa coupling to a Higgs field. In
this case, the standard decoupling theorem of Appelquist and Carazzone
\ref\ac{T. Appelquist and J. Carazzone, \prd{11}{1975}{2856}.} does not apply,
since a heavy fermion necessitates a large coupling. D'Hoker and Farhi found
that a Wess-Zumino term \ref\wz{J. Wess and B. Zumino, \plb{37}{1971}{95}.} is
naturally generated in the process of integrating out the heavy field.
Consequently, no anomaly problems arise in the low energy theory even when the
heavy fermion is involved in an anomaly cancellation mechanism in the full
theory. In fact, it seems that as far as perturbation theory is concerned, all
of the low energy structure of the full theory can be reproduced in the
effective theory.

More recently, Banks and Dabholkar \ref\bd{T. Banks and A. Dabholkar,
\prd{46}{1992}{4016}.} have suggested that there is no way to match the
nonperturbative behavior of the original theory using a local low energy
Lagrangian for the light fields. In particular, they suggest a problem with
the 't~Hooft instanton process \ref\thooft{G. 't~Hooft, \prl{37}{1976}{8}.}.
In the original theory, a light fermion cannot decay via such a process
because the difference between light and heavy fermion number is a conserved
quantity. In another language, the 't~Hooft instanton has zero modes for all
fermions, so that all generations must participate in any instanton induced
interaction. On the other hand, in the low energy effective theory, they find
nothing to prevent a light fermion from decaying through instantons.

In this paper, we examine this apparent paradox. The key observation is that
the contribution to the {\it effective Lagrangian in the full theory} from
integrating out the heavy fermions is a non-analytic function of the Higgs
field. But the {\it Lagrangian of the effective low energy
theory}\foot{Note the two very different uses of the word ``effective.''}
must be a polynomial. This is not a failure of decoupling, but it does mean
that an extra step is required in matching onto the low energy theory. We must
expand the full theory effective Lagrangian in powers of the shifted Higgs
field and truncate the expansion at some finite order. We resolve the paradox
by showing that such a consistent matching from the full theory to the low
energy effective theory described by a local, polynomial Lagrangian generates
terms which forbid the instanton configurations in the low energy theory. The
instantons do not exist because the would-be instanton
configurations have infinite action.

First, note that there are two ways to have a separation of scales between
light and heavy fermions. In one scenario, assumed in the analysis in \df, the
light fermions are either below or at the VEV scale, while the heavy fermions
are much heavier than the VEV. The gauge bosons are also light, and the
physical Higgs potential is massively fine-tuned to keep the Higgs mass at the
VEV scale. There are several problems with this approach. First, the theory is
strongly coupled, and we don't really know how to deal with it in perturbation
theory. In addition, we expect solitons to exist at the VEV scale which will
carry the heavy fermion number \ref\dfs{E. D'Hoker and E. Farhi,
\plb{134}{1984}{86}; \npb{241}{1984}{109}.}. Therefore, our supposed low
energy effective theory will not even have the right particle content.
Finally, there are indications (e.g., \ref\eg{M. Einhorn and G. Goldberg,
\prl{57}{1986}{2115}.}) that a theory with a large fermion mass may well be
inconsistent because the Yukawa coupling increases at small distances and may
blow up at a finite scale. It makes no sense to give a fermion a mass greater
than the cutoff scale of the theory.

For these reasons, we
will concentrate on an alternative scenario, proposed by D. Kaplan and
discussed in \bd. Here, the heavy fermion is below the VEV scale, while
everything else is much lighter still. A problem suggested in \bd, apart from
the fact that the Higgs mass has to be fine tuned, is that the Higgs effective
potential is unbounded below for large values of the field. In fact, in the
regime that we are considering, the turning point predicted by the one-loop
effective potential may not be very far from the VEV. However, we know that
our ``full'' theory is in fact an effective theory because, for example,
it is not
asymptotically free. We may imagine, for example, that the Higgs doublet is a
composite
formed by an ultrafermion condensate not far above the $SU(2)\times
U(1)$ breaking scale \ref\kg{D. Kaplan, H. Georgi, and S. Dimopoulos,
\plb{136}{1984}{187}.}. Then the minimum in the Higgs effective potential
which
occurs at the VEV is the true minimum, rather than a metastable vacuum which
it appears to be if we assume the original theory to be valid at all scales.
Of course, we must then be prepared to consider nonrenormalizable terms in our
``full'' theory. Because the ultracolor physics (or what ever it is) does not
break the $SU(2)\times U(1)$ symmetry, we can ignore it in the discussion
below.

    We now show that the instanton which appears to allow the light fermion to
decay in the effective theory in fact does not exist. Every non zero winding
number configuration is associated with
divergent Lagrangian density near the core.
The high density
arises from terms in the effective theory obtained by
integrating out heavy fermion loops with multiple Higgs insertions. These
non-renormalizable terms involve large numbers of derivatives of the angular
part of the Higgs field.
After integration of the Lagrangian density over space-time (where the
integral has been regularized
near the origin in some reasonable way), we obtain an infinite action.
Therefore the configuration does not contribute to the path integral. This is
not surprising, because we expect on physical grounds that high frequency
configurations should not contribute in the low energy theory.

We imagine computing the Lagrangian for the low energy theory
(to first order in the loop expansion) by integrating out heavy fermion
loops
in the manner of \df, except that the heavy fermion Yukawa coupling is
less than
unity, and the ratio of scales is finite. In the low energy Lagrangian, there
will be a tower of nonrenormalizable interactions
with dimension greater than four. In practice, we must always truncate the
low energy Lagrangian at some point and ignore the interactions with dimension
greater than some integer, $n>4$. This is a perfectly natural procedure which
we might implement automatically without thinking too deeply about it if we
were computing the low energy Lagrangian in perturbation theory about the
broken symmetry vacuum state. However, this simple point implies a dramatic
difference between the effective theory and the full theory.
It implies that in the effective theory we must
truncate the expansion of the Lagrangian in powers of derivatives, powers of
the gauge
field, and powers of the {\it shifted} physical Higgs field.
It is important to note in particular that we are not allowed to sum
up the powers of the shifted Higgs field. If we were to do so, we
would obtain terms involving only the unshifted
Higgs doublet (and its derivatives), but they would be non-analytic in this
field. The fact
that, at least to one-loop order, one can write down a (non-analytic)
effective
Lagrangian in the
full theory which is a function of the light fermion fields, the gauge
fields, and the Higgs doublet field, is not relevant to the
problem of constructing the Lagrangian in the low energy theory,
where we have no choice but to separate the Higgs doublet into its radial
and angular components, shift the physical field, and construct a
(truncated) expansion in the shifted field.

Let us understand exactly what types of terms arise in our effective
theory, after we integrate out the heavy fermions. We are concerned with
diagrams which have one heavy fermion loop. These generate terms involving
the gauge field $A_{\mu}$ and the spinless doublet field
$\Phi=\phi U$ (with $U$ the unitary part and $\phi=\phi'+v$ the
magnitude --- $\phi'$ is the physical Higgs field). For simplicity,
we will consider the situation in which
the two components of the heavy fermion doublet are degenerate.
Then the field $\Phi$ always appears multiplied by a single Yukawa coupling,
which we will call $\lambda$.\foot{If the two components of the doublet were
not degenerate, the coupling would be a matrix in flavor space and the
expressions below would be more complicated.} We expand these expressions in
inverse powers of the VEV $v$. Using the field variables $A_{\mu}$,
$\phi'$, and $U$, we obtain terms of the form
\eqn\terms{{(A_{\mu})^p\phi'^q(\pd A_{\mu})^r(\pd \phi')^s(\pd U)^t\over
v^{p+q+2r+2s+t-4}}\,,}
where $(\pd U)^t$ is a generic symbol for $t$ derivatives acting on $U$ or
$U^{\dg}$, multiplied by appropriate factors of $U$ and $U^{\dg}$. For
example,
$(\pd U)^4$ may stand for $(U\pd_{\mu} U^{\dg})^4$ or $\pd_{\mu} \pd^{\mu}
\pd_{\nu} U \pd^{\nu} U^{\dg}$, etc. Similarly, each of the other factors
in \terms\ stands generically for any one of a number of possible
expressions. The trace of the terms in \terms\ must of course be
taken in the Lagrangian density.

Although many of these terms may potentially give an infinite contribution
to the instanton action, in the following discussion we will concentrate on
isolating the $(\pd U)^4/v^0$ terms. Since $U$ has
constant magnitude, its derivative near the center of the instanton behaves
as $1/r\,$; hence these terms behave as $1/r^4$ and yield a logarithmically
divergent contribution to the action. Of course, terms with higher powers
of $\pd U$ yield even more severe divergences.
The $(\pd U)^4/v^0$ terms arise from
diagrams
with one heavy fermion loop with a total of either two or four insertions of
the $\Phi$ and $\Phi^{\dg}$ fields.
A very important point in these diagrams is the modification of the heavy
fermion propagator from ${\not\! k/k^2}$ to
\eqn\chiral{
 {\not\! k+\lambda\Phi P_++\lambda\Phi^\dagger P_-\over
k^2-\lambda^2\phi^2}\,,}
where $P_\pm=(1\pm\gamma_5)/2$. This modification arises
from insertions, on the fermion propagators in the diagram, of vertices with
zero-momentum $\Phi$ or $\Phi^{\dg}$ fields. The infinite geometric series of
such terms yields the modified
propagator. The inclusion of diagrams with
zero-momentum insertions is necessary
because diagrams with more than four external fields produce terms
with four derivatives; their contribution to the four-derivative term is
obtained by setting all but four momenta in the diagram
equal to zero. The chiral structure of \chiral\
requires, e.g., an odd number of zero-momentum
insertions between a $\Phi$ and a $\Phi$, hence the three possible forms of the
modified
propagator. As mentioned above, the presence of $\phi$ fields in
the denominator will be seen to be of the utmost importance.

The diagram with, e.g., four alternating $\Phi$ and $\Phi^{\dg}$ insertions
will yield various expressions, including ones of the form
\eqn\pp{\int\!\! dy\lambda^4\phi^4{P_{\sss(4)}\over
(\lambda^2\phi^2-P_{\sss(2)})^2}} and
\eqn\ppp{\int\!\! dy\lambda^4\phi^4\ln\left({\lambda^2\phi^2-P_{\sss(2)}\over
\mu^2}\right),}
where $\mu$ is a renormalization scale.
Here $P_{\sss(2)}$ and $P_{\sss(4)}$ are generic symbols for second and
fourth
order polynomials
in the $\phi$ momenta $p_i$ and the $U$ or $U^{\dg}$ momenta $q_i$
$(i=1,\ldots,4)$.
These polynomials are symmetric in
$p$ and $q$, as well as in $i$ (with $\sum_1^4 (p_i+q_i)=0$), and their
coefficients depend on the Feynman parameters $y$.
If we were to perform the integration over Feynman parameters, we would obtain
expressions which are non-analytic in $P'_{\sss(2)}/\lambda^2\phi^2$,
where $P'_{\sss(2)}$ now stands for a second order polynomial in the
external momenta with fixed coefficients. However, for convenience we will
delay doing the parameter integrals.

We now must decide what to do with
non-analytic expressions such as \pp\ and \ppp, since we want to obtain a
local
low energy
Lagrangian, i.e., a polynomial in the light fields (in particular
$\phi'=\phi-v$) and derivatives.

Let us concentrate on \pp.
In order to get the $\phi$'s out of the
denominator we
must perform two expansions: ({\rm i}) we must expand in momenta, and ({\rm
ii})
we
must expand in the physical Higgs field $\phi'$.

Consider the momentum expansion. If the momenta are small, such
that
$P_{\sss(2)}<\lambda^2\phi^2$, \pp\ becomes
\eqn\pexp{\int\! dy P_{\sss(4)}
\Bigl(1+{2P_{\sss(2)}\over\lambda^2\phi^2}+
{3P_{\sss(2)}^2\over\lambda^4\phi^4}
+\cdots\Bigr)\,.}
The Feynman parameter integrals may now easily be done,
to yield an expansion with fixed coefficients.
The leading term in \pexp\ produces the $(\pd U)^4/v^0$ term
that we wish to concentrate on. In position
space this leading term yields expressions such as
\eqn\first{
{[\partial_{\mu}(\phi U^{\dagger})\partial^{\mu}(\phi U)]^2\over\phi^4}
\qquad {\rm and}\qquad
{[\partial_{\mu}(\phi U^{\dagger})\partial_{\nu}(\phi U)]^2\over\phi^4}
\,.}
Expanding the first gives
\eqn\three{{1\over
\phi^4}[(\partial_{\mu}U^{\dagger}\partial^{\mu}U)^2\phi^4+
2(\partial_{\mu}U^{\dagger}\partial^{\mu}U)(\partial_{\nu}\phi)^2\phi^2+
(\partial_{\mu}\phi)^4]\,.}
The first term in \three,
\eqn\us{(\partial_{\mu}U^{\dagger}\partial^{\mu}U)^2\,,}
is the one we are interested in; it has $1/r^4$ behavior at the origin
and leads to infinite instanton action.

The above expansion is clearly the correct one for perturbatively computing
(low energy) amplitudes in the effective theory. Now consider nonperturbative
effects in the low energy theory. For any
non-zero winding number configuration,
there is a region near the core where our expansion breaks down
because $P'_{\sss(2)}/\lambda^2\phi^2$ becomes large. However,
when we examine the question of instantons in the low energy
theory, we should do so without referring back to the full theory from which
it originally
came.
{}From the point of view of the effective theory, we have a low energy
Lagrangian which we must use to evaluate the action for any configuration
of light fields. It is not relevant that large
$P'_{\sss(2)}/\lambda^2\phi^2$ is outside of the radius of convergence
of the original expansion. What is relevant is that the low energy
Lagrangian which we have obtained matches correctly both the perturbative
and the nonperturbative physics which is predicted by the Lagrangian
of the full theory.

There is nothing very surprising about the fact that winding configurations
do not contribute to the effective theory physics.
The Goldstone boson fields are rapidly varying in an instanton configuration
--- their derivatives approach infinity at the center. In the full theory
instanton, the Goldstone boson fields are a part of an $SU(2)$ multiplet with
the physical Higgs field, and this field vanishes at the center of the
instanton. Thus, in the full theory, the rapid
variation of the Goldstone boson fields is just a coordinate
singularity; nothing physical is actually getting large. But in the low
energy theory, the connection between the Higgs field and the Goldstone boson
fields is severed by the matching corrections discussed above. In the low
energy theory, the singularity in the Goldstone boson derivatives at the
center of the instanton is real. Thus the instanton configuration is rapidly
varying and should be irrelevant to the physics of the low energy theory.

We have determined that no configuration with non-zero instanton
number exists in the effective theory, so the question of light fermion
decay is resolved. What about the zero-instanton sector? In the full
theory,
a pair of instanton and anti-instanton can exist, with
the amplitude suppressed by $e^{-N\!M\!r}$ for separations larger
than ${\cal O} (1/M)$ ($N$ is the number of heavy fermion zero modes, and
$M$ is the heavy fermion mass) \bd. From the statistical mechanics
point of view, instantons are confined by a linear potential for
separations larger than ${\cal O} (1/M)$. In the semiclassical approximation,
however, the widely separated pairs do not contribute to any amplitudes,
because no such configurations are (even approximate) stationary points
of the action. On the other hand, small instantons (with size smaller
than ${\cal O} (1/M)$) can be separated by a short enough distance so that
the potential is not linear; thus approximate stationary points may exist
and contribute to light fermion scattering amplitudes.
Of course,
in the effective theory, the amplitude for a process where light fermions
are destroyed by the instanton and created by the anti-instanton is
exactly zero, regardless of the size of the instanton or anti-instanton.
Therefore, there
is no discrepancy between the two theories for large instantons. As far
as small instanton pairs are concerned, their effects can be matched
onto the low energy Lagrangian at second order in the instanton density
expansion using the framework developed in
\ref\sam{H. Georgi and S. Osofsky, Harvard University Preprint HUTP-93/A021.}.
In any case, nonperturbative effects in the zero-instanton
sector are not a matter of great concern. They will contribute only to
processes which can already take place perturbatively in the low energy
theory, and they will be exponentially suppressed in the small gauge
coupling constant.

The possible existence of nontopological (bag-like) solitons is another
matter which deserves attention \bd. These are
three-dimensional configurations where
the Higgs field bends towards zero in a finite region of space, allowing
a ``light'' heavy fermion to be located at the center.
The same situation arises as with topological solitons.
These bag-like objects would be interesting (and problematic) if we
could make the
heavy fermion much heavier than the VEV (see the fourth paragraph of this
paper).
In the more realistic scenario with which we are dealing,
balancing between the second order and higher order derivative terms
will make the mass of any potentially stable soliton be on the order
of the VEV, and hence heavier than the heavy fermion itself.
Thus, there are no states
in the low energy theory which carry the heavy fermion quantum number.
In the full theory, it seems clear that such nontopological objects of all
sizes are unstable.

It has been argued \ref\steve{S. Hsu, Harvard University Preprint
HUTP-93/A019.} that the apparent impossibility of
correctly computing the effective action for all light field
configurations using a low
energy Lagrangian is a problem for the theory, because knowing the
effective action for all configurations is equivalent to knowing
all one-particle-irreducible Green's functions.
Furthermore, it was asserted that only the S-matrix can be matched in
the effective theory.
In fact,
there seems to be no obstacle to matching low momentum one-light-particle
irreducible Green's functions in the full and effective theories, and we
do not expect to be able to match high momentum Green's functions in
a low energy theory.
More concretely, in our model there is no discrepancy in contributions
to bosonic Green's functions coming from the one instanton sector, since
instantons do not contribute to the bosonic effective action in either
the full or the effective theory. As far as small instanton pairs in the
full theory are concerned, they may be important for computing short
distance bosonic correlators, but these are precisely the correlators
which we do not compute in a low energy theory.

Normally, in the presence of instantons, the matching process requires
a consistent expansion in both the number of loops and the instanton
density \sam. In our case, this approach is not directly applicable because
the
one-loop correction to the instanton action in the effective theory is not
small. A self-consistent approach here seems
to be as follows. First, we perform
the matching calculation to one loop order, as shown above for the heavy
fermion loops. At this order, we already see that instantons have infinite
action and thus do not
contribute to physical quantities in the effective theory.
We then go on to do the matching at higher
orders in the loop expansion, noticing that the instanton action in the
low energy theory remains infinite at each order in this expansion.
Also, since
processes involving (small) instanton dipoles are relevant in the full theory,
we also expand in powers of the dipole density in the full theory, with the
effects
being matched onto high dimension terms in the low energy Lagrangian.

\vfill\eject

\listrefs

\bye